\def\bar{\overline}

\def\k{\kappa}
\def\e{\epsilon}

\def\bar{\overline}
\def\l{\lambda}

\documentstyle [12pt,epsf]{article}
\begin{document}
\baselineskip=22 pt
\setcounter{page}{1}
\thispagestyle{empty}
\topskip 2.5  cm
\setcounter{page}{1}
\thispagestyle{empty}
\vspace{1 cm}
\centerline{\Large\bf Leptogenesis in  Neutrino Textures with Two Zeros}
\vskip 1.5 cm
\centerline{{\large \bf Satoru Kaneko}
\renewcommand{\thefootnote}{\fnsymbol{footnote}}
\footnote[1]{E-mail address: kaneko@muse.sc.niigata-u.ac.jp}\ ,
 \qquad {\large \bf Makoto Katsumata}
\renewcommand{\thefootnote}{\fnsymbol{footnote}}
\footnote[2]{E-mail address: katumata@muse.sc.niigata-u.ac.jp} }
\vskip 0.5 cm 
\centerline{\large\bf and}
\vskip 0.5 cm 
\centerline {{\large \bf Morimitsu Tanimoto}
\renewcommand{\thefootnote}{\fnsymbol{footnote}}
\footnote[3]{E-mail address: tanimoto@muse.sc.niigata-u.ac.jp}
 }
\vskip 0.8 cm
 \centerline{ \it{Department of Physics, Niigata University, 
 Ikarashi 2-8050, 950-2181 Niigata, JAPAN}}
\vskip 2 cm
\centerline{\bf ABSTRACT}\par

\vskip 1 cm

The leptogenesis is studied in the neutrino textures with two  zeros,
  which  reduce the number of independent phases of the  CP violation. 
The phenomenological favored neutrino textures with two zeros
are decomposed into the Dirac neutrino mass matrix and
the right-handed Majorana one in the see-saw mechanism.
Putting the condition to suppress the $\mu \rightarrow e\gamma$ decay 
 enough, the texture zeros of the Dirac neutrino mass matrix are fixed 
in the framework of the MSSM  with right-handed neutrinos.  
These textures have only one CP violatig phase.  
The magnitude of each entry of the Dirac mass matrix is
determined in order to explain the baryon asymmetry of the universe
 by  solving the Boltzman equations.
The relation between the leptogenesis and the low energy CP violation 
  is presented in these  textures. 

\newpage
\topskip 0. cm
\section{Introduction}
In these years empirical understanding of the mass and mixing of 
 neutrinos have been advanced \cite{SKam,SKamsolar,SNO}.
The KamLAND experiment selected the neutrino
mixing solution that is responsible for the solar neutrino problem
nearly uniquely \cite{KamLAND}, only large mixing angle solution.
We have now good understanding concerning the neutrino
mass difference squared ($\Delta m^2_{\rm atm}$, $\Delta m^2_{\rm sun}$) and 
flavor mixings of neutrinos ($\sin^2 2\theta_{\rm atm}$,
 $\tan^2 \theta_{\rm sun}$) \cite{Lisi}.

 The texture zeros of the neutino mass matrix have been discussed
 to explain these data  \cite{Obara}.
Recently, Frampton, Glashow and Marfatia \cite{Fram} found 
acceptable textures of the neutrino mass matrix with two independent
vanishing entries in the basis with the diagonal charged lepton mass matrix.
They  have been examined in details phenomenologically
\cite{Xing1,Xing2,Barbieri,HKT}. 
These textures can be
 decomposed into the Dirac and the right-handed Majorana neutrino mass 
matrix with zeros  in the see-saw mechanism \cite{Seesaw}.
 The texture zeros of neutrinos reduce the number of independent phases of 
the CP violation 
\footnote{
In general, the independent CP violating phases are six
for  three generations   without a left-handed Majorana mass term.}.

In this paper, we study the leptogenesis \cite{Fu} based on these textures. 
The leptogenesis is an important  candidate to explain
 the observed baryon asymmetry of the universe.
The CP violation required for the leptogenesis stems from
phases in the right-handed  sector,  whereas
the CP violation in  the neutrino oscillations \cite{CP} can be described 
by the phase in the left-handed neutrino mixing matrix \cite{MNS}.
Therefore, once the texture of the Dirac neutrino mass matrix is given,
 the leptogenesis links with 
the low energy CP violation  in the neutrino oscillations
 \cite{lepto,MSSM,leptcp,lepttexture,FGY,moro}.

We have decomposed the acceptable textures of the neutrino mass matrix
 \cite{Fram} into the Dirac neutrino mass matrix and
the right-handed Majorana one \cite{decomp}.
Putting the condition to suppress the $\mu \rightarrow e\gamma$ decay enough, we fix the  textures of the Dirac neutrino mass matrix,
 which  have  only one CP violating phase.  
 The magnitude of each entry of the Dirac mass matrix is
determined in order to explain the baryon asymmetry of the universe
 by  solving the Boltzman equations.

 We have already discussed  the textures of the Dirac neutrinos
 only for the special case
\footnote{This case  corresponds to
 $m=4$, $n=2$, $x=2$, $y=2$, $z=2$ and  $w=1$ in this paper.} 
in the previous work \cite{KT}, in which 
the rough approximate dilution factor  was used.
  In this paper,  we  fix the texture in the framework of the minimal 
supersymmetric standard model (MSSM)  with right-handed neutrinos (RN)
by solving the Boltzman equations  numerically.
The relation between the leptogenesis and the low energy CP violation 
  is presented  in these textures.

 The texture zeros of the Dirac neutrino mass matrix are presented
 in section 2.  The CP violating phases and  magnitude of 
  matrix elements are discussed in section 3.
 In section 4, the CP asymmetry in the leptogenesis and 
the low energy CP violation are discussed  for the selected  textures.
Numerical results are given in section 5. 
Section 6 is devoted to the summary.

\section{Texture Zeros of Dirac Neutrino Mass Matrix}
 Let us start with decomposing  the textures presented 
 by Frampton, Glashow and Marfatia \cite{Fram}
  into the Dirac neutrino mass matrix and the
 right-handed Majorana neutrino one \cite{decomp}.

 There are seven acceptable textures with two independent zeros
for the effective neutrino mass matrix   $M_{\nu}$ \cite{Fram}.
Among them, the textures $\rm A_1$ and $\rm A_2$ in  ref.\cite{Fram}, which  
correspond to the hierarchical neutrino mass spectrum, are strongly favored 
by the phenomenological analyses \cite{Xing1,Xing2,Barbieri,HKT}.  Therefore,  we discuss these two  textures
in this paper.
The textures of the $\rm A_1$ and $\rm A_2$ types are written :

\begin{equation}
{\rm A_1}:  {\bf M_{\nu}}=
\left ( \matrix{
{\bf 0} & {\bf 0} & \times \cr
{\bf 0} & \times & \times \cr
\times & \times & \times \cr} \right ) \simeq 
   m_0 \left( \matrix{  0  &  0 &  \l \cr  0  & 1  &  1      \cr 
                 \l & 1  &  1 \cr }  \right) ,  \ \ 
{\rm A_2}: 
 {\bf M_{\nu}}=
\left ( \matrix{
{\bf 0} & \times & {\bf 0} \cr
\times & \times & \times \cr
{\bf 0} & \times & \times \cr} \right ) \simeq
 m_0 \left( \matrix{  0  &  \l &  0 \cr  \l   & 1  &  1      \cr 
                 0 & 1  &  1 \cr }  \right)  ,
\label{texture}
\end{equation}
\noindent  with  $\l\simeq 0.22$ in the basis of the diagonal  charged lepton 
mass matrix, and $m_0$ is the scale of the neutrino mass.

 In principle these textures are given at the low energy scale because
 the experimental data are put to determine zeros, however, the structure
of zeros in the mass matrix is not changed by the one-loop 
renormalization group  equations of the MSSM  \cite{RGE}.
Therefore, we discuss the see-saw realization in these textures
 at the right-handed Majorana neutrino ($N_1,\ N_2,\ N_3$) scale:
\begin{equation}
\bf M_\nu=  m_D\   M_R^{\rm -1}\  m_D^{\rm T} \ ,
\end{equation} 
\noindent where  $\bf m_D$ and  $\bf M_R$ are the mass matrices
for the Dirac neutrino masses   and the right-handed Majorana neutrino ones,
 respectively.
As far as we exclude the possibility that  zeros are 
originated from cancellations among coefficients in the see-saw mechanism, 
the see-saw realization of these seven textures are not trivial.
Then, these zeros should come from zeros of the Dirac neutrino mass matrix 
and the right-handed Majorana neutrino one.
 These results are summarized in the ref.\cite{decomp}.

Once the right-handed Majorana neutrino mass matrix is specified,
the Dirac neutrino ones are selected to reproduce the textures 
in eq.(\ref{texture}). In order to study the leptogenesis, the 
diagonal basis of the  right-handed Majorana neutrino mass matrix is favored. 
However, there is no solution for acceptable  textures unless we 
consider cancellations between matrix elements of the Dirac and Majorana ones.
Therefore, we take a following  texture
 for the right-handed Majorana neutrinos with three independent parameters:
\begin{equation}
{\bf M_R} =\left ( \matrix{  {\bf 0}  & \times &  {\bf 0}\cr
 \times &   \times &  {\bf 0}\cr {\bf 0} & {\bf 0} & \times \cr} \right ) \ ,
\label{R}
\end{equation}
\noindent 
where $\times's$ denote non-zero entries.
It should be noted that specifying the texture of the right-handed Majorana 
neutrinos is a choice of weak basis.
Furthermore we can take the matrix  in eq.(\ref{R}) to be real.  Then,
the CP violating phases appear only in the Dirac neutrino mass matrix.
In this case, we have six Dirac neutrino mass matrices, which reproduce
  the  texture  $\rm A_1$ in eq.(\ref{texture}):
\begin{eqnarray}
{\bf m_D} =&&\left ( \matrix{   {\bf 0} & \times &  {\bf 0}\cr
  {\bf 0} & \times &  \times\cr  \times & \times  & \times \cr} \right ) \ ,
\quad 
\left ( \matrix{  {\bf 0} & \times &    {\bf 0}\cr
  {\bf 0} & \times &  \times\cr  \times  &{\bf 0} &  \times \cr} \right ) \ ,
\quad 
\left ( \matrix{ {\bf 0} &\times &   {\bf 0}\cr
  {\bf 0} & \times &   \times\cr  \times & \times  & {\bf 0} \cr} \right ) \ ,
\nonumber \\
&&
\left ( \matrix{   {\bf 0} & \times &   {\bf 0}\cr
  {\bf 0} & \times &   \times\cr  \times & {\bf 0} & {\bf 0}\cr} \right ) \ ,
\quad
\left ( \matrix{  {\bf 0} & \times &   {\bf 0}\cr
  {\bf 0} &  {\bf 0} &  \times\cr  \times & \times  & \times \cr} \right ) \ ,
\quad
\left ( \matrix{   {\bf 0} & \times &    {\bf 0}\cr
  {\bf 0} & {\bf 0} & \times\cr \times  & {\bf 0} & \times \cr} \right ) \ ,
\label{A11}
\end{eqnarray}
 \noindent
where $\times's$ denote  complex numbers.
 For the texture $\rm A_2$ in eq.(\ref{texture}), we obtain
\begin{eqnarray}
{\bf m_D} =&&\left ( \matrix{  {\bf 0} & \times &    {\bf 0}\cr
 \times &   \times &  \times\cr   {\bf 0}  & \times &\times \cr} \right ) \ ,
\quad 
\left ( \matrix{  {\bf 0} & \times &    {\bf 0}\cr
 \times &   \times &  \times\cr  {\bf 0} & {\bf 0} & \times \cr} \right ) \ ,
\quad 
\left ( \matrix{ {\bf 0} & \times &     {\bf 0}\cr
 \times &  \times &{\bf 0} \cr   {\bf 0} & \times & \times \cr} \right ) \ ,
\nonumber\\
&&\left ( \matrix{  {\bf 0} &\times &    {\bf 0}\cr
  \times &  {\bf 0} &\times\cr  {\bf 0}  & \times & \times \cr} \right ) \ ,
\quad
\left ( \matrix{  {\bf 0} & \times &    {\bf 0}\cr
  \times & {\bf 0} & \times\cr {\bf 0} & {\bf 0}  & \times \cr} \right ) \ ,
\quad
\left ( \matrix{  {\bf 0} & \times &   {\bf 0}\cr
 \times & {\bf 0} & {\bf 0}\cr  {\bf 0}&\times & \times \cr}\right ) \  .
\label{A21}
\end{eqnarray}
 
 We can select the texture by the lepton flavor violation (LFV).  
Many authors have studied the LFV in the MSSM+RN  assuming the relevant 
neutrino  mass matrix \cite{Borz,LFV1,LFV2,Sato}.
In the MSSM with soft breaking terms, 
there exist  lepton flavor violating terms such as 
off-diagonal elements of slepton mass matrices 
and trilinear couplings (A-term).
It is noticed that large neutrino Yukawa couplings and large lepton mixings 
 generate the large LFV in the left-handed slepton masses. For example, 
the decay rate of  $\mu\rightarrow e \gamma$ can be approximated as follows:
\begin{eqnarray}
{\rm \Gamma} (\mu\rightarrow e \gamma) \simeq 
\frac{e^2}{16 \pi} m^5_{\mu} F
 \left| \frac{(6+2 a_0^2)m_{S0}^2}{16 \pi^2}
({\bf Y_\nu  Y_\nu^\dagger})_{21} \ln\frac{M_X}{M_R}  \right|^2\ ,
\label{rate}
\end{eqnarray}
\noindent where the neutrino Yukawa coupling matrix  $\bf Y_\nu$ 
 is given as  ${\bf Y_\nu}={\bf m_D}/v_2$ ($v_2$ is a VEV of Higgs)
at the right-handed mass scale $M_R$, and 
  $F$ is a function of masses and mixings for SUSY particles.  
In eq.(\ref{rate}),
 we assume the universal scalar mass $(m_{S0})$ for all scalars and
the universal A-term $(A_f=a_0 m_{S0} Y_f)$ at the GUT scale $M_X$. 
Therefore  the branching ratio $\mu \rightarrow e \gamma$ depends 
considerably on the texture \cite{Sato}.

Many works have shown that this branching ratio is too large 
\cite{Sato}.
However, zeros in the Dirac neutrino mass matrix may suppress enough
the branching ratio.  The condition is that 
$({\bf m_D m_D^\dagger})_{21}$ and
 $({\bf m_D m_D^\dagger})_{31}\times ({\bf m_D m_D^\dagger})_{32}$ are tiny
compared with  other ones.
Therefore we put  the following conditions to select the texture zeros:
\begin{eqnarray}
({\bf m_D m_D^\dagger})_{21}=0 \ , \qquad
({\bf m_D m_D^\dagger})_{31}\times ({\bf m_D m_D^\dagger})_{32}=0 \ .
\label{condlfv}
\end{eqnarray} 
 \noindent Then, the branching ratio of  $\mu\rightarrow e \gamma$ is 
safely predicted to be below the present experimental
 upper bound $1.2\times 10^{-11}$ \cite{exp} due to the texture zeros.
Actually, the case of the texture  zeros  was examined
carefully in ref. \cite{Ellis}.

\section{CP Violating Phases in  Dirac Neutrino Mass Matrix}
Key ingredients of the leptogenesis are the structure of CP violating phases
and the magnitude of each entry for the Dirac neutrino mass matrices 
in  eq.(\ref{A11})  
since the right-handed Majorana neutrino mass matrix is taken to be real.
 Although the non-zero entries $\times$ in the Dirac neutrino mass matrix
are  complex, three phases are removed 
by the re-definition of the left-handed neutrino fields.
There is no freedom of re-definition for the right-handed ones
in the basis with  real $\bf M_R$.
Furthermore, we move to the  diagonal basis of the right-handed Majorana 
neutrino mass matrix in order to calculate the magnitude of the leptogenesis.
Then,  the Dirac neutrino mass matrices $\bf \bar m_D$ in the new basis
is given  as follows:
\begin{equation}
  {\bf \bar m_D} =  {\bf P_L \  m_D \ O_R} \ ,
\end{equation}
\noindent
where $\bf P_L$ is a diagonal phase matrix and $\bf O_R$ is the 
orthogonal matrix which diagonalizes $\bf M_R$ as  $\bf O_R^T M_R O_R$.
Since  the phase matrix $\bf P_L$ can remove one phase in each row of
 $\bf m_D$, three phases disappear  in ${\bf \bar m_D}$.

 By taking three eigenvalues of $\bf M_R$  as follows 
\footnote{The minus sign of $M_1$ is necessary to reproduce 
$\bf M_{R}$.  This minus sign is transfered to  
$\bf  m_D$ by the right-handed diagonal phase matrix $diag(i,1,1)$.}:
\begin{equation}
  M_1=-\l^m  M_0  \  , \qquad M_2= \l^n  M_0  \ , \qquad  M_3=  M_0  \ ,  
\end{equation}
\noindent  where $m$ and $n$ are positive integer
with $m>n$, 
 we obtain the orthogonal matrix $\bf O_R$ as 
\begin{equation}
  {\bf O_R} =  \left ( \matrix{ \cos{\theta} &  \sin{\theta}&  0\cr 
 - \sin{\theta} &  \cos{\theta} & 0 \cr  0 &  0  & 1 } \right )  \ ,
\qquad\quad  \tan^2 {\theta}= \l^{m-n} \ .
\end{equation}

For the texture  $A_1$, only one texture satisfies  conditions in 
eq.(\ref{condlfv}):
\begin{eqnarray}
   {\rm A_1^D}\ :\ 
   {\bf m_{D}} =m_{D0}
   \left( 
       \matrix{  
                 { 0}  & a \l^{x} &  { 0} \cr
                 { 0}  & {0}  &  b       \cr 
                 c\l^{z} e^{i\rho} & { 0}  &  f  \cr }  \right) \ ,
\label{textureA1}
\end{eqnarray}
where $a$, $b$, $c$, $f$ are the real order one coefficients, 
$x$ and $z$ are positive integer,
 which should satisfy the following conditions:
\begin{eqnarray}
     x+z = \frac{m+n}{2}+1\ \ ,\quad 
     2z \geq m\ \ ,\quad 
     m>n \ ,
\end{eqnarray}
\noindent in order to reproduce the  texture  $A_1$ in eq.(\ref{texture}). 
These conditions  lead to the inequality:
 \begin{eqnarray}
 \frac{m}{2} \leq   z \leq     \frac{n+m}{2}+ 1 < m+1 \ .
\end{eqnarray}
\noindent Therefore, we obtain  sets for $(m,\ z)$,
where $m$ starts from 3 since the hierarchy is assumed for the
right-handed Majorana masses and $m+n$ is even, as follows:
\begin{eqnarray}
   (m=3,\ \ z=2,3) \ , \quad    (m=4,\  z=2,3,4)\ ,\quad   (m=5,\ \ z=3,4,5)\ ,  \ \ \ \cdots \ .
\end{eqnarray} 
\noindent We will discuss the best choice of  $m$ and $z$ in the leptogenesis.

 For the texture $A_2$, we find two textures of the Dirac neutrino mass 
 matrix, which satisfy   conditions in eq.(\ref{condlfv}).
The first one is
\begin{eqnarray}
   {\rm A^D_2(1)}\ :\ 
   {\bf m_{D}} =m_{D0}
   \left( 
       \matrix{   { 0}  &  a \l^{x} &  { 0} \cr
                 b\l^{y}  &  { 0}  &  c e^{i\rho} \cr 
                 { 0}  &  { 0}  &  f  \cr
              } 
   \right) \ ,
\label{textureA2}
\end{eqnarray}
\noindent where
 $m$, $n$, $x$ and  $y$ satisfy the following conditions:
\begin{eqnarray}
     x+y=\frac{m+n}{2}+1\ \ ,\ \ 
     2y \geq m\ \ ,\ \ 
     m>n \ .
\end{eqnarray}
\noindent  The second one is
\begin{eqnarray}
   {\rm A^D_2(2)}\ :\ 
   {\bf m_{D}} =m_{D0}
   \left(\matrix{  
                 {0}  &  a \l^{x} &  { 0} \cr
                 b\l^{y}  &  { 0}  &  {0} \cr 
                 {0}  &  d\l^{w}e^{i\rho}  &  f  \cr } \right) \ ,
\label{textureA20}
\end{eqnarray}
\noindent where $m$, $n$, $x$, $y$ and $w$ satisfy the following conditions:
\begin{eqnarray}
     x+y=\frac{m+n}{2}+1\ \ ,\ \ 
     y = \frac{m}{2}\ \ ,\ \ 
     y+w = \frac{m+n}{2}\ \ ,\ \ 
     m>n \ ,
\end{eqnarray}
\noindent which lead to
\begin{eqnarray}
     x=1+\frac{n}{2}\ \ ,\ \ 
     y = \frac{m}{2}\ \ ,\ \ 
     w = \frac{n}{2}\ \ ,\ \ 
     m>n \ .
\label{cond3}
\end{eqnarray}
\noindent
 Thus, $x$, $y$ and $w$ are fixed if $m$ and $n$ are given 
in this case.

\section{Leptogenesis and Low Energy CP Violation}
 Once the textures are fixed, we can discuss the leptogenesis
 numerically.
 Summing up the one-loop vertex and self-energy corrections,
the lepton number
asymmetry (CP asymmetry) for the lightest heavy Majorana neutrino ($N_1$) 
decays into $l^{\mp} \phi^{\pm}$  in the MSSM+RN\cite{MSSM} is given by

\begin{eqnarray}
&& \epsilon_1 =
\frac{\Gamma_1-\overline{\Gamma_1}}{\Gamma_1+\overline{\Gamma_1}} =  
-\frac{1}{8\pi v_2^2}\frac{1}{({\bf \bar m_D}^{\dagger} {\bf \bar m_D})_{11}}
 \sum_j {\rm Im}[({\bf \bar m_D}^{\dagger}   {\bf\bar m_D})_{1j}^2]
 \ \left [f\left (\frac{M_j^2}{M_1^2}\right ) + g\left (\frac{M_j^2}{M_1^2}\right ) \right ] \ , \nonumber\\
&&  f(x)=\frac{2\sqrt{x}}{x-1} \ ,
 \qquad\quad  g(x)=\sqrt{x}\ \ln \left(\frac{1+x}{x} \right )  \ ,
\label{epsilon} 
\end{eqnarray}
\noindent where  $M_1 < M_2,\ M_3$ is assumed, and 
$v_2=v \sin\beta$ with $v=174$GeV.
In our analyses, we take $\sin\beta\simeq 1$.
The lepton asymmetry $Y_L$ is related to the CP asymmetry through 
the relation:
\begin{equation} 
Y_L=\frac{n_L-n_{\bar{L}}}{s}=\k\,\frac{\epsilon_1}{g_{\ast}}\ ,
\end{equation}
where $s$ denotes the entropy density, 
$g_{\ast}$ is the effective number of relativistic degrees of freedom
contributing to the entropy and $\kappa$ is 
the so-called dilution factor which
accounts for the washout processes (inverse decay and lepton number violating
scattering). 
In the case of the MSSM+RN, one gets $g_{\ast}=232.5$.

The produced lepton asymmetry $Y_L$ is converted into a net baryon asymmetry
$Y_B$ through the $(B+L)$-violating sphaleron processes. One  finds the
relation \cite{harvey}
\begin{equation} 
Y_B=\xi \ Y_{B-L}=\frac{\xi}{\xi-1} \ Y_L \  , \qquad
\xi= \frac{8\ N_f + 4\ N_H}{22\ N_f + 13\ N_H} \ ,
\end{equation}
where $N_f$ and $N_H$ are the number of fermion families and complex Higgs
doublets, respectively. Taking into account  $N_f=3$ and $N_H=2$ in the MSSM, 
we get 
\begin{equation}  
Y_B = -\frac{5}{18}\,Y_L \ .
\end{equation}


On the other hand, the low energy CP violation, which is a measurable 
quantity in the long baseline neutrino oscillations \cite{CP}, 
is given by the Jarlskog determinant $J_{\scriptscriptstyle CP}$ \cite{JCP}, 
which is calculated by
\begin{equation}  
   det [{\bf  M_\ell M_\ell^\dagger,  M_\nu M_\nu^\dagger}]
= -2 i   J_{\scriptscriptstyle CP}  
(m_\tau^2-m_\mu^2) (m_\mu^2-m_e^2)(m_e^2-m_\tau^2)
 (m_3^2-m_2^2) (m_2^2-m_1^2)(m_1^2-m_3^2)  ,
\end{equation}
\noindent where $\bf M_\ell$ is the diagonal charged lepton mass matrix, 
and $m_1$, $m_2$, $m_3$ are neutrino masses.

 It is very interesting to investigate links
between the leptogenesis ($\epsilon_1$) and the low energy CP violation
 ($J_{\scriptscriptstyle CP}$)  in each  texture.
Let us begin with investigating the case $A_1^D$ in eq.(\ref{textureA1}).
In this texture, we get 
\begin{eqnarray}
     \e_1 \simeq  -\frac{3m_{D0}^2}{8\pi v^2_2}
          \frac{\l^m + \l^n}{a^2 \l^{m+2x} + c^2\l^{n+2z}}
          \frac{c^2 f^2 \l^{m+2z}}{1+\l^{m-n}} \sin2\rho
        =   -\frac{3m_{D0}^2}{8\pi v^2_2} \frac{c^2 f^2 \l^{m+2z}}
                  {a^2 \l^{2m-2z+2} + c^2\l^{2z}} \sin2\rho \   ,
\label{eA1}
\end{eqnarray}
\begin{eqnarray}
        J_{\scriptscriptstyle CP} = \frac{1}{64}
\frac{\Delta m^2_{\rm atm}}{\Delta m^2_{\rm sun}}\ 
                   a^2 b^4 c^4 f^2 \l^{-2m-n+2x+4z}\sin2\rho
        =   \frac{1}{64}
\frac{\Delta m^2_{\rm atm}}{\Delta m^2_{\rm sun}}\ 
 a^2 b^4 c^4 f^2 \l^{-m+2z+2}\sin2\rho  \  ,
\label{JA1}
\end{eqnarray}
\noindent
where we used experimental values 
$\Delta m^2_{\rm atm}$  and  $\Delta m^2_{\rm sun}$
assuming  $m_1\ll  m_2\ll m_3$ 
\footnote{ As is well known,  the CP violation vanishes in the neutrino oscillation in the case of $\Delta m^2_{\rm sun}=0$, which corresponds to
 the $\l=0$ limit in our case.}.
 The value of $m_{D0}^2$ is estimated from  $m_0 M_0$.
The $\e_1$ is classified by $z$.
The first  term of the denominator suppressed only in the case of  $z=m/2$.
Then,  we have
\begin{eqnarray}
  \e_1  &\simeq& -\frac{3m_{D0}^2}{8\pi v_2^2} f^2 \l^m\sin2\rho \ , 
\qquad (m=2z) \ , \nonumber \\
      J_{\scriptscriptstyle CP} &=&
                  \frac{1}{64}
\frac{\Delta m^2_{\rm atm}}{\Delta m^2_{\rm sun}}\ a^2 b^4 c^4 f^2 \l^{2}
  \sin2\rho \ .
\end{eqnarray}
\noindent
For other $z$'s $(m/2<z<m+1)$, the first  term  dominated the 
 the denominator in  eq.(\ref{eA1}).  Then, we get
\begin{eqnarray}
     \e_1 &\simeq&  -\frac{3m_{D0}^2}{8\pi v_2^2}
                 \frac{c^2 f^2}{a^2}\l^{-m+4z-2}\sin2\rho \ , \qquad
 (\frac{m}{2}<z< m+1) \nonumber\\
 J_{\scriptscriptstyle CP} &=&
 \frac{1}{64}
\frac{\Delta m^2_{\rm atm}}{\Delta m^2_{\rm sun}}\ 
 a^2 b^4 c^4 f^2 \l^{-m+2z+2}\sin2\rho \ ,
\end{eqnarray}
\noindent where $ J_{\scriptscriptstyle CP}$ is smaller than $\l^2$.

The next case is given in the texture $A^D_2(1)$.
The result is given by replacing $z$ with $y$ as follows:
\begin{eqnarray}
    && \e_1 \simeq  \frac{3m_{D0}^2}{8\pi v_2^2}
            \frac{\l^m + \l^n}{a^2 \l^{m+2x} + b^2\l^{n+2y}}
            \frac{b^2 c^2 \l^{m+2y}}{1+\l^{m-n}} \sin2\rho
        =   \frac{3m_{D0}^2}{8\pi v_2^2}
             \frac{b^2 c^2 \l^{m+2y}}
                  {a^2 \l^{2m-2y+2} + b^2\l^{2y}}
             \sin2\rho \ ,  \nonumber\\
   \label{Sig2}\\
       && J_{\scriptscriptstyle CP}=
                   \frac{1}{64}
\frac{\Delta m^2_{\rm atm}}{\Delta m^2_{\rm sun}}\  
      a^2 b^4 c^2 f^4 \l^{-2m-n+2x+4y}\sin2\rho
= \frac{1}{64} \frac{\Delta m^2_{\rm atm}}{\Delta m^2_{\rm sun}}
  \ a^2 b^4 c^2 f^4 \l^{-m+2y+2}\sin2\rho .
\nonumber
\end{eqnarray}
\noindent Therefore, the numerical result is the same as the case in $A_1^D$
except for the relative sign.

By putting eq.(\ref{cond3}), we get for the  case of the texture $A^D_2(2)$
\begin{eqnarray}
     \e_1 &\simeq &  -\frac{3m_{D0}^2}{8\pi v_2^2}
          \frac{\l^m + \l^n}{a^2 \l^{m+2x} + b^2\l^{n+2y} + d^2\l^{m+2w}}
          \frac{d^2 f^2 \l^{2m+2w}}{\l^m+\l^{n}}
          \sin2\rho
\nonumber\\ 
        &\simeq& -\frac{3m_{D0}^2}{8\pi v_2^2}
          \frac{d^2 f^2}{b^2 + d^2}
          \l^m\sin2\rho \ ,
          \nonumber\\
  J_{\scriptscriptstyle CP}&=& \frac{1}{64}
 \frac{\Delta m^2_{\rm atm}}{\Delta m^2_{\rm sun}}\ 
                    a^2 b^6 d^2 f^2 \l^{-3m-2n+2x+6y+2w}\sin2\rho
  \nonumber\\
            &=& \frac{1}{64}
\frac{\Delta m^2_{\rm atm}}{\Delta m^2_{\rm sun}}\ 
                   a^2 b^6 d^2 f^2 \l^{2}\sin2\rho       \ .
\label{Sig3}
\end{eqnarray}
\noindent
This case is the simplest one since $\e_1$ is given by only  $m$ and 
the magnitude of $J_{\scriptscriptstyle CP}$ is $\l^2$.

\section{Numerical Results}
 In order to calculate the baryon asymmetry, 
we need the dilution factor involves the integration of the 
 Boltzmann equations \cite{pu}.

The Boltzmann equations for the $N_1$ number densities and the
$N_{B-L}$ asymmetry is given as:
\begin{eqnarray}
&&\frac{dN_1}{dz}=-\left( \frac{\Gamma_D}{H z}+
 \frac{\Gamma_S}{H z}\right )(N_1-N_1^{eq}) \ , \nonumber \\
 \nonumber \\
&&\frac{dN_{B-L}}{dz}= -\e_1  \frac{\Gamma_D}{H z} 
(N_1-N_1^{eq})-\frac{\Gamma_W}{H z} N_{B-L} \ ,\quad\quad z=\frac{M_1}{T} \ , 
\end{eqnarray}
where $H$  is Hubble parameter at $z=1$, and 
$\Gamma_D$, $\Gamma_S$ and $\Gamma_W$ account for the decay and inverse decay
process, the scattering processes,  the total washout processes,
respectively.
In our numerical calculation, we take into only the process involving
the interaction with the top quark for $\Delta L = 1$ scattering processes
 $\Gamma_S$ since this process is dominant one.

In Fig.1, we show the evolution of neutrino number density $Y_{N_1}$, 
$Y_{N_1}^{eq}$ and the lepton asymmetry  $|Y_{L}|$ for 
the typical case of the texture $A_1^D$.
The equilibrium distribution for $N_1$ is represented by a dashed line
 against $z$.
 In this calculation,   
$\sin 2\rho=1/3$, $M_0=10^{15}$GeV, $m=6$, $n=2$, $z=3$ are taken,
   and the coefficient $f$ is $1$. These parameters correspond to
 $M_1\simeq 10^{11}$GeV.
The gray area shows the measured value for the lepton asymmetry,
which is derived from  the new data of WMAP \cite{WMAP}:
\begin{equation}
\eta = 6.5 ^{+0.4}_{-0.3} \times 10^{-10} \ ( 1\ \sigma) \ .
\end{equation}
\noindent This value  corresponds to 
\begin{equation}
Y_B = (7.9\sim 11) \times 10^{-11} \ ( 3\ \sigma) \ .
\end{equation}
The generated asymmetry is consistent with the observed one as seen 
in Fig.1.
\begin{figure}
\epsfxsize=15.0 cm
\centerline{\epsfbox{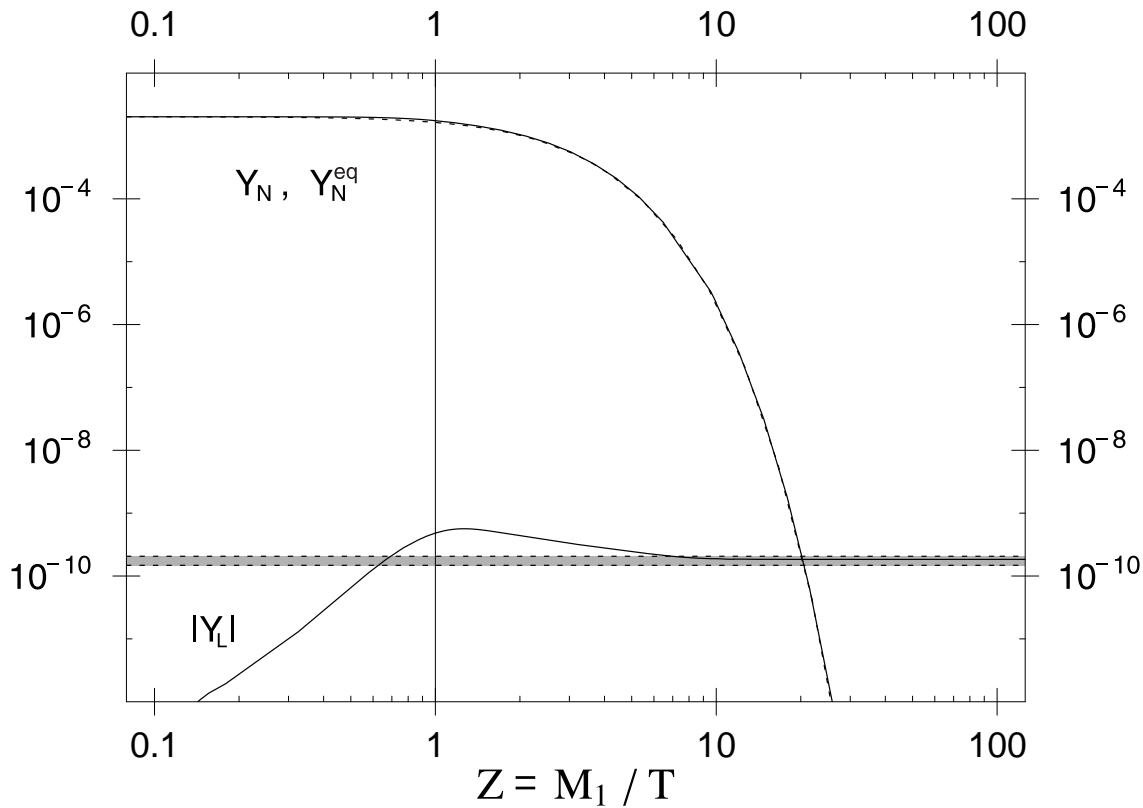}}
\caption{Time evolution of neutrino number density $Y_{N_1}$, 
and lepton asymmetry  $|Y_{L}|$ for the texture $A_1^D$,
 where $M_0=10^{15}$GeV, $\sin 2\rho=1/3$, $m=6$, $n=2$, $z=3$ are taken.
 The equilibrium distribution $Y_{N_1}^{eq}$ is represented by a dashed line
 against $z=M_1/T$. The gray area shows the measured value for the lepton 
asymmetry.}
\end{figure}

In Fig.2, we show the relation of the predicted $J_{\scriptscriptstyle CP}$ 
and the baryon asymmetry $|Y_B|$ of the $A_1^D$ texture in the case of 
$M_0=10^{15}$GeV, where 13 points are predictions  for
 different $m$ and $z$.
The relative sign between $J_{\scriptscriptstyle CP}$ and  $Y_B$  is minus
 as seen in eq.(\ref{eA1}) and eq.(\ref{JA1}).
The prediction in the case of  $(m=7,\ z=4)$  is  consistent with 
the observed baryon asymmetry taking  $\sin 2\rho=1$,
 while  $J_{\scriptscriptstyle CP}$ is $0.007$.
  The cases of $(m=5,\ z=3)$ and $(m=6,\ z=3)$ give rather large
 asymmetry. In these cases, the prediction could be consistent with
the observed baryon asymmetry by taking smaller $\sin 2\rho$.
 Then, $J_{\scriptscriptstyle CP}$ is at most $0.01$.
It is  noticed that  the ambiguity of factor $2\sim 3$ should be
 taken into  predictions since coefficients  $a$, $b$, $c$, $d$ and $f$ 
are taken to be 1.
\begin{figure}
\epsfxsize=15.0 cm
\centerline{\epsfbox{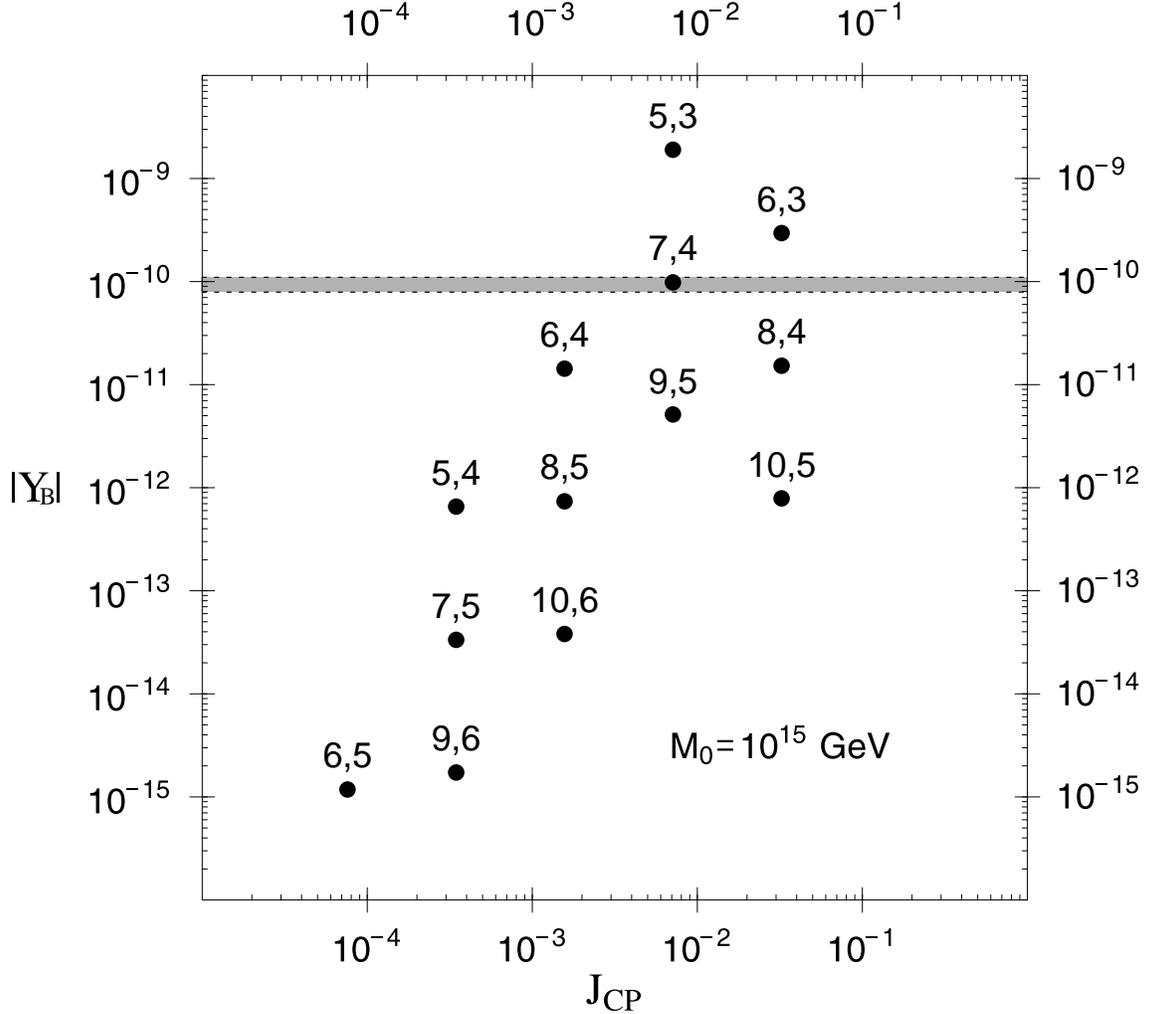}}
\caption{ Relation of the predicted $J_{\scriptscriptstyle CP}$ and
the baryon asymmetry $|Y_B|$ of the texture $A_1^D$  in the case of 
$M_0=10^{15}$GeV with  $\sin 2\rho=1$, 
where numbers above black points denote ($m$,$z$),respectively.}
\end{figure}

 Fig.3 corresponds to the case of $M_0=10^{14}$GeV.
The prediction in the cases of $(m=5,\ z=3)$
is consistent with the observed baryon asymmetry
while  $J_{\scriptscriptstyle CP}$ is around  $0.007$.
The cases of $(m=4,\ z=3)$ and $(m=6,\ z=3)$ may be consistent
  if we take  account of   the ambiguity  of order one  coefficients.
 In the case of $(m=4,\ z=2)$, the prediction could be consistent with
the observed baryon asymmetry by taking smaller $\sin 2\rho$.
 
 The case of  $M_0=10^{13}$GeV is shown in Fig.4.
The prediction in the case of $(m=4,\ z=2)$
 is consistent with the observed asymmetry.
The predicted value of $J_{\scriptscriptstyle CP}$ is  $0.03$.
Since the predicted value may be multiplied by a factor $2\sim 3$
due to the order one factor  $c^2 f^2/a^2$,  the case of  
$(m=5,\ z=3)$ could be consistent with the observed asymmetry.
The case of $(m=3,\ z=2)$ is also  consistent with
the observed baryon asymmetry by taking smaller $\sin 2\rho$.

\begin{figure}
\epsfxsize=15.0 cm
\centerline{\epsfbox{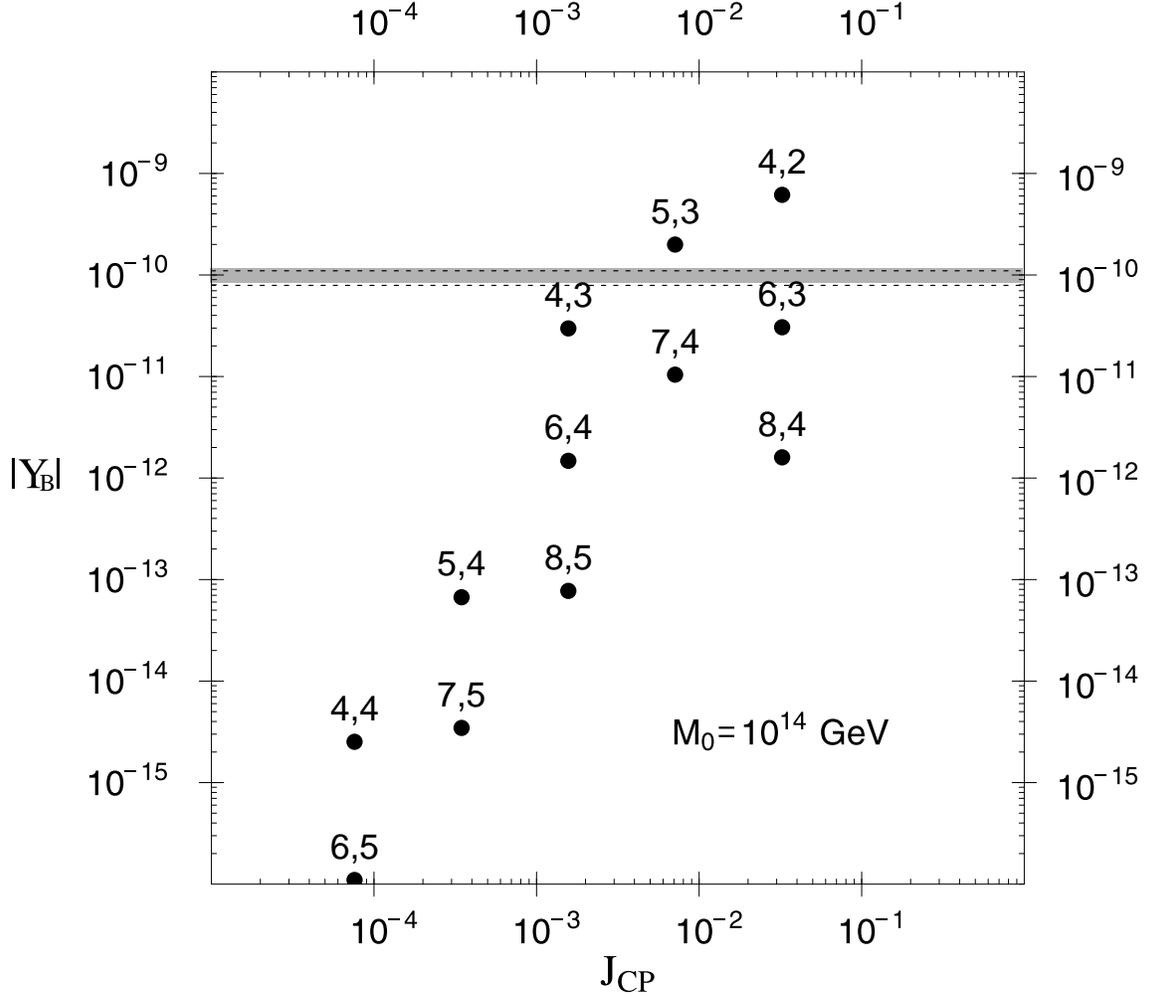}}
\caption{ Relation of the predicted $J_{\scriptscriptstyle CP}$ and
 the baryon asymmetry $|Y_B|$ of the texture $A_1^D$  in the case of 
 $M_0=10^{14}$GeV with   $\sin 2\rho=1$,
 where numbers above black points denote  ($m$,$z$), respectively.}
\end{figure}
\begin{figure}
\epsfxsize=15.0 cm
\centerline{\epsfbox{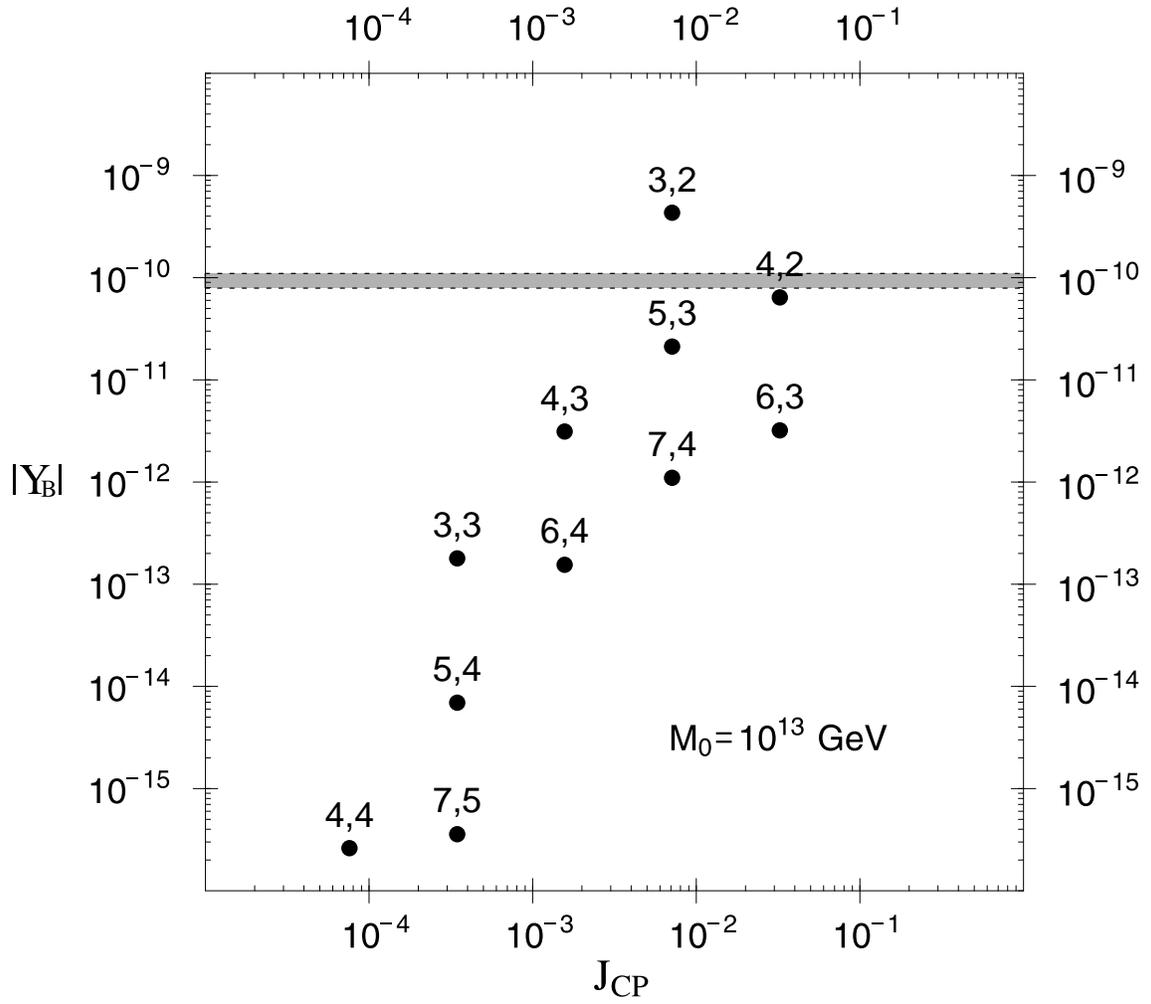}}
\caption{ Relation of the predicted $J_{\scriptscriptstyle CP}$ and
 the baryon asymmetry $|Y_B|$ of the texture $A_1^D$  in the case of 
 $M_0=10^{13}$GeV with   $\sin 2\rho=1$,
 where numbers above black points denote  ($m$, $z$), respectively.}
\end{figure}

In Fig.5, predictions of the baryon asymmetry $|Y_B|$ are
presented  against $M_1$ 
 for  $M_0=10^{13},\ 10^{14},\ 10^{15}$GeV.
It is found that  $M_1=10^{10}\sim 10^{11}$GeV is consistent with 
the observed asymmetry.
This result is consistent with the lower bound of $M_1$ in ref.\cite{DI}.
\begin{figure}
\epsfxsize=10.0 cm
\centerline{\epsfbox{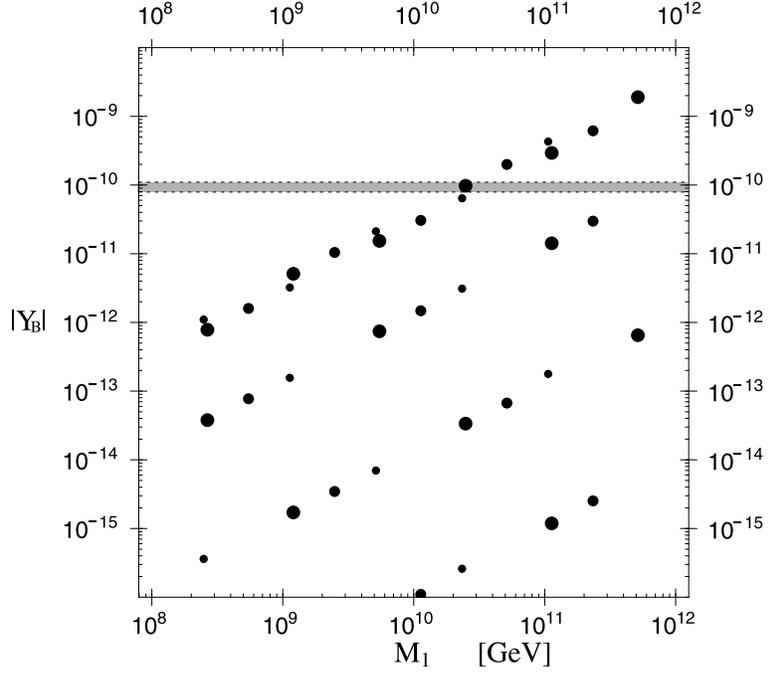}}
\caption{Baryon asymmetry $|Y_B|$ versus $M_1$ in the texture $A_1^D$
with   $\sin 2\rho=1$.
 Big, middle, small black points correspond
 to  $M_0=10^{15}$, $10^{14}$, $10^{13}$GeV, respectively.}
\end{figure}
\begin{figure}
\epsfxsize=10.0 cm
\centerline{\epsfbox{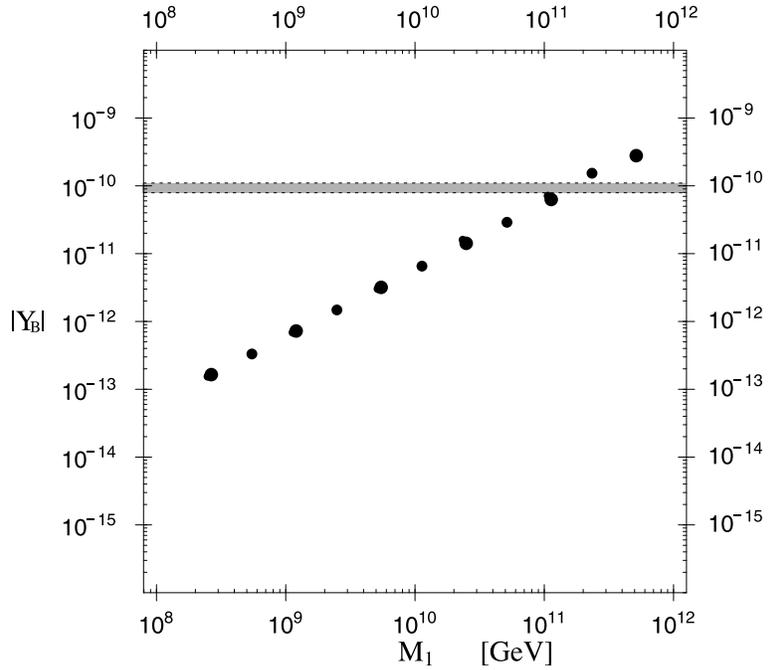}}
\caption{Baryon asymmetry $|Y_B|$ versus $M_1$ in the texture $A^D_2(2)$
 for different $m$ and $M_0$ by taking $\sin 2\rho=1$. 
Big, middle, small black points correspond 
to  $M_0=10^{15}$, $10^{14}$, $10^{13}$GeV, respectively.
  Many predicted points overlap.}
\end{figure}
 In the case of the texture  $A^D_2(1)$, results are the same 
as the ones in the texture  $A^D_1$ apart from
the relative sign between   $J_{\scriptscriptstyle CP}$ and $Y_B$,
which is plus as seen in eq.(\ref{Sig3}).
On the other hand,  the case of the texture $A^D_2(2)$ texture is simple
because the prediction depends only on $m$ and $M_0$.
 We show the $|Y_B|$ versus $M_1$ in  Fig.6 for
 $M_0=10^{13},\ 10^{14},\ 10^{15}$GeV, while  $J_{\scriptscriptstyle CP}$ 
 is $\l^2$ being independent of $m$.
 It is found that predictions almost depend on only $M_1$.
It is found that 
$M_1=5\times 10^{10}\sim 5\times 10^{11}$GeV is consistent with 
the observed asymmetry.  The typical Dirac neutrino mass matrices  
are given by $m=6$ for $M_0=10^{15}$GeV,  
$m=4$ for $M_0=10^{14}$GeV and  $m=3$ for $M_0=10^{13}$GeV.

\section{Summary}
The leptogenesis is studied in the texture zeros of neutrinos,  which 
reduce the number of independent phases of the  CP violation. 
The phenomenological favored neutrino textures with two zeros
are decomposed into the Dirac neutrino mass matrix and
the right-handed Majorana one in the see-saw mechanism. Putting the 
condition to suppress the $\mu \rightarrow e\gamma$ decay enough,
 the  Dirac neutrino textures are fixed  in the framework of the MSSM+RN.  
These textures have only one CP violating phase.  
The magnitude of each entry of the matrix is
determined in order to explain the baryon asymmetry of the universe
 by  solving the Boltzman equations. The preferred Dirac mass matrix
depends on $M_0$. We have typical ones,
 $(m=7,\ z=4)$ for $M_0=10^{15}$GeV, $(m=5,\ z=3)$ for
 $M_0=10^{14}$GeV and $(m=4,\ z=2)$ for $M_0=10^{13}$GeV.
The relation between the baryon asymmetry and 
 $J_{\scriptscriptstyle CP}$ has been  discussed in these textures. 
 For the textures  $A_1^D$ and  $A^D_2(2)$, 
 the relative sign  between the  baryon asymmetry and 
 $J_{\scriptscriptstyle CP}$ is plus, 
while  it is minus for the texture $A^D_2(1)$.
Thus, it is very important to observe the relative sign
to distinguish  textures.
It is also noticed  that $M_1=10^{10}\sim 5\times 10^{11}$GeV is required
to reproduce  the observed asymmetry, and
 $J_{\scriptscriptstyle CP}$ is expected to be $0.007\sim 0.03$.
We expect the observation of  $J_{\scriptscriptstyle CP}$ 
in the future experiments.

\vskip 0.3  cm
 
This research is supported by the Grant-in-Aid for Science Research,
 Ministry of Education, Science and Culture, Japan (No.12047220). 

\newpage

\end{document}